\documentclass[pmlr]{jmlr}

\usepackage{longtable}

\usepackage{booktabs}
\usepackage[load-configurations=version-1]{siunitx} 

\theorembodyfont{\upshape}
\theoremheaderfont{\scshape}
\theorempostheader{:}
\theoremsep{\newline}

\jmlryear{2023}
\jmlrworkshop{Causal Analysis Workshop Series (CAWS)}

\title[py-tetrad and rpy-tetrad]{Py-Tetrad and RPy-Tetrad: A New Python Interface with R Support for Tetrad Causal Search \titletag{}}

\author{\Name{Joseph D. Ramsey\nametag{}} \Email{jdramsey@andrew.cmu.edu}\\
\addr Department of Philosophy, Carnegie Mellon University, Pittsburgh, PA
\AND
\Name{Bryan Andrews} \Email{andr1017@umn.edu}\\
\addr Department of Psychiatry \& Behavioral Sciences, University of Minnesota, Minneapolis, MN}

\editor{Editor's name}

\begin{document}

\maketitle

\begin{abstract}
We give novel Python and R interfaces for the (Java) Tetrad project for causal modeling, search, and estimation. The Tetrad project is a mainstay in the literature, having been under consistent development for over 30 years. Some of its algorithms are now classics, like PC and FCI; others are recent developments. It is increasingly the case, however, that researchers need to access the underlying Java code from Python or R. Existing methods for doing this are inadequate. We provide new, up-to-date methods using the JPype Python-Java interface and the Reticulate Python-R interface, directly solving these issues. With the addition of some simple tools and the provision of working examples for both Python and R, using JPype and Reticulate to interface Python and R with Tetrad is straightforward and intuitive.
\end{abstract}
\begin{keywords}
Tetrad, Python, R, JPype, Reticulate, causal search
\end{keywords}

\section{Introduction}
\label{sec:intro}

Tetrad \citep{ramsey2018tetrad}\footnote{\href{https://github.com/cmu-phil/tetrad}{https://github.com/cmu-phil/tetrad}} is a well-established package for causal modeling, search, and estimation that has been continuously developed since the early 1990s. Not only is it the source for several now-classic algorithms, such as the celebrated Peter/Clark (PC) \citep{spirtes2000causation} and Fast Causal Inference (FCI) \citep{spirtes2000causation} algorithms, but it contains implementations of more recent state-of-the-art algorithms as well. Algorithms in Tetrad include, but are not limited to:
\begin{itemize}
    \item Fast Greedy Equivalence Search (FGES) \citep{ramsey2017million}, an implementation of the Greedy Equivalent Search (GES) \citep{chickering2002optimal} algorithm;
    \item Build Pure Clusters (BPC) \citep{silva2006learning}, an algorithm for searching over latent variable structures using tests of variable tetrads;
    \item Greedy Relaxations of Sparsest Permutation (GRaSP) \citep{lam2022greedy}, a permutation-based algorithm that provides high-accuracy output for dense graphs.
\end{itemize}

However, it is increasingly the case that users need to access the underlying code of Tetrad. This can present an obstacle to those unfamiliar with Java since most data scientists now primarily use Python \citep{vanrossum2010python} or R \citep{rref}. Python is beneficial for machine learning and is also increasingly used in the sciences. At the same time, R has been a mainstay of scientific and statistical research for many years. Python and R are scripting languages that can mock up ideas quickly, wrangle data, generate plots, do statistical analyses, etc. In addition, Python has become a go-to language for scientific algorithmic development, so a great deal of software has become readily available for general use in Python that is not readily accessible from Java. 

The Python project, JPype \citep{nelson2020jpype}, provides a ready solution for accessing the underlying code of Tetrad from Python; it allows easy access, using Python-native syntax, to any Java class, method, or field, so we have created a package, py-tetrad,\footnote{\href{https://github.com/cmu-phil/py-tetrad}{https://github.com/cmu-phil/py-tetrad}} to show how to use JPype to interact with Tetrad Java code. Py-tetrad provides three things. First, it provides simple tools for translating datasets and graphs between Python and Tetrad. Second, it provides a class, TetradSearch, that handles everyday operations without needing to resort to JPype programming explicitly. And third, it provides numerous examples of using TetradSearch and JPype to interface Python with Tetrad. 

To access Tetrad from R, we take advantage of the Reticulate R package,\footnote{\href{https://rstudio.github.io/reticulate/}{https://rstudio.github.io/reticulate/}} which provides indirect access to Tetrad through py-tetrad. We call the resulting project rpy-tetrad; it is located in a subdirectory of the py-tetrad project.\footnote{\href{https://github.com/cmu-phil/py-tetrad/tree/main/pytetrad/R}{https://github.com/cmu-phil/py-tetrad/tree/main/pytetrad/R}} R through Reticulate can share data and graphs directly with py-tetrad, and py-tetrad can then translate them to Java. Routing the connection from R to Java through Python has the advantage of fewer ``moving parts''; updates to py-tetrad are immediately accessible to R, so the projects do not need to be updated separately. As with py-tetrad, rpy-tetrad provides numerous examples.

Installation instructions for py-tetrad and rpy-tetrad may be found in the ReadMe files on their respective GitHub pages, and example files for use may also be found in these directories on GitHub. These instructions will be kept up to date if changes to Java, Tetrad, or Python require it. The example files in py-tetrad and rpy-tetrad will also be kept up to date as changes to Java, Tetrad, or Python happen.

It is worth noting that the idea of interfacing Python and R with Tetrad is not new; previous work had produced the packages py-causal (for Python) and r-causal (for R) using the Causal Command tool\footnote{\href{https://github.com/bd2kccd/causal-cmd}{https://github.com/bd2kccd/causal-cmd}} (see the comparison in Table \ref{comparisontable}). For those using py-causal or r-causal, we recommend transitioning to py-tetrad and rpy-tetrad for two reasons. First, the Java versions used in py-causal and r-causal are now outdated and no longer receive updates. Using py-tetrad and rpy-tetrad, users can benefit from the significant improvements made to Tetrad in recent years. Second, JPype enables access to the entire Tetrad codebase from Python, not just a select portion. While this may not be essential for users interested in specific methods already supported by py-causal or r-causal, it dramatically facilitates exploring aspects of Tetrad that are not directly supported.\footnote{We have had many complaints about the use of javabridge for py-causal, especially on Macs, where it can be almost impossible to install. We have also had complaints about the use of rjava for r-causal in its handling of data transfers. JPype, so far, appears to be bug-free and is easy to install, which is yet another reason why we switched to it.}

\begin{table}[ht]
\begin{center}
\begin{tabular}{|c|c|cc|c|} 
\hline
 & py-causal & r-causal & py-tetrad & rpy-tetrad \\
\hline
Tetrad Version & 6.8.0 & 6.8.0 & 7.5.0+ & 7.5.0+ \\
Last Developed & 2017 & 2017 & 2023+ & 2023+ \\
Tetrad Access & javabridge & rjava & JPype & JPype \\
Python Version & 2.7+ & - & 3.5+ & 3.5+ \\
R Version & - & 3.3.0+ & - & 4+ \\
\hline
\end{tabular}
\caption{\label{comparisontable}Comparison of py-causal, r-causal, py-tetrad, rpy-tetrad}
\end{center}
\end{table}

\section{Prepackaged Tools}
\label{sec:functions}

Prepackaged methods are provided in the `tools' directory of the project\footnote{\href{https://github.com/cmu-phil/py-tetrad/tree/main/pytetrad/tools}{https://github.com/cmu-phil/py-tetrad/tree/main/pytetrad/tools}} to translate datasets between Tetrad and Python and to translate graphs from Tetrad to Python. A class, TetradSearch,\footnote{\href{https://github.com/cmu-phil/py-tetrad/blob/main/pytetrad/tools/TetradSearch.py}{https://github.com/cmu-phil/py-tetrad/blob/main/pytetrad/tools/TetradSearch.py}} is provided to give access to Tetrad without using JPype calls; this class encapsulates a wide swath of Tetrad functionality and is accessible from R. These are not necessarily the only tools that will eventually be provided, but they are helpful for a wide variety of tasks.

The dataset translators serve two purposes. First, they provide translations of data between Tetrad and Python in a way that makes them useful for most purposes. They are simple, fast, and effective; even large datasets are translated quickly. They can translate data that is continuous, discrete, or a mixture of continuous and discrete columns. From Python to Tetrad, discrete columns are detected by the column type; these column types can be set directly in the data frame if the Python data loader does not already set them to the appropriate type. Second, Python has an abundance of tools for data preprocessing and wrangling that researchers can now directly incorporate into their pipelines.

Several graph translation methods from Tetrad to Python are provided to handle a variety of purposes. Users can, of course, write their own graph translators for special purposes if the provided formats are inadequate. One can do this by wrapping one of these methods and extending its functionality.

A graph can be retrieved from Tetrad in the following formats:

\begin{enumerate}
    \item As a GeneralGraph object in causal-learn. The causal-learn package\footnote{\href{https://causal-learn.readthedocs.io/en/latest/index.html}{https://causal-learn.readthedocs.io/en/latest/index.html}} supports a graph object in Python compatible with Tetrad's EdgeListGraph; the translation is direct, and the returned graph can be manipulated directly in Python.
    \item As an edge matrix in PCALG \citep{kalisch2012causal}. This is an edge matrix in which tail, arrow, star, circle, and null (no endpoint) endpoints are represented as distinct integers, which is compatible with Tetrad's EdgeListGraph.
    \item As a simple DOT string. This is the form of a graph used by Graphviz \citep{ellson2002graphviz} to plot graphs. Graphviz is a sophisticated graph plotting tool with many options; our DOT format gives a basic DOT output. For more nuanced DOT outputs, the user can implement a similar method in Python (as we do for our Python example below).
    \item As a model specification for the lavaan \citep{rosseel2017package} R package. Any directed and acyclic graph can be saved in this format.
\end{enumerate}

If one uses JPype directly, the entire codebase of Tetrad is at one's fingertips; code examples are given in the py-tetrad project to show how to do this, but we will give an example below to show how powerful this can be. Of course, when accessing the Tetrad codebase, it helps to know the various classes, methods, and fields that one can use in one's JPype code. For this, it helps to have the documentation for the Tetrad package available for reference. This is given in the usual Java format as a set of Javadocs,\footnote{The version of Javadocs at the time of writing of this report is at \href{https://www.phil.cmu.edu/tetrad-javadocs/7.4.0/}{https://www.phil.cmu.edu/tetrad-javadocs/7.4.0/}} which can be accessed online, though one should check for updates;\footnote{\href{https://github.com/cmu-phil/py-tetrad}{https://github.com/cmu-phil/py-tetrad}} the current Tetrad version for these is 7.4.0, though these are updated as revisions to Tetrad are made.

It should be noted as well that JPype allows for Java interfaces to be implemented in Python, so it is possible to define tests or scores in Python for use in Java search methods. An example of this is provided in the py-tetrad package.\footnote{\href{https://github.com/cmu-phil/py-tetrad/blob/main/pytetrad/general_scoring_example.py}{https://github.com/cmu-phil/py-tetrad/blob/main/pytetrad/general\_scoring\_example.py}.}

\section{A Code Example in Python}
\label{sec:example-python}

A typical py-tetrad workflow starts by loading a dataset, wrangling the data, and then optionally plotting histograms and scatter plots. Python provides robust tools for these tasks, so there is no need to rely on Tetrad for these steps. One then converts the data into a Tetrad dataset object before passing it to a Tetrad search procedure. The procedure returns a graph in Tetrad format, which can optionally be converted back to Python for further processing, such as calculating statistics in a simulation study or aggregating multiple runs in a bootstrapped study; examples of both are available in py-tetrad. 

We give an example of a bootstrapped study below. Tetrad has built-in bootstrapping facilities, which we will showcase in our subsequent R example for comparison, but we will do the bootstrapping in Python to show how it can be done. The bootstrapped study uses the Apple Watch Fitbit data \citep{fuller2020using}, which is available in easily loaded format at the indicated location.\footnote{\href{https://github.com/cmu-phil/example-causal-datasets/tree/main/real/apple-watch-fitbit}{https://github.com/cmu-phil/example-causal-datasets/tree/main/real/apple-watch-fitbit}} This code example is included in the py-tetrad repository.\footnote{\href{https://github.com/cmu-phil/py-tetrad/blob/main/pytetrad/jpype_example.py}{https://github.com/cmu-phil/py-tetrad/blob/main/pytetrad/jpype\_example.py}}

First, we load the data and make a knowledge object. This example has a mixture of continuous and discrete columns, so we need to correctly set the column types for our variables so they will be translated to Tetrad properly. We also use Tetrad's knowledge facility to put variables into ``knowledge tiers'' to ensure that variables in later tiers cannot cause variables in earlier tiers. We use the Degenerate Gaussian mixed-type score \citep{andrews2019learning} with the SP-FCI algorithm (see \cite{raskutti2018learning} and \cite{ogarrio2016hybrid}).\footnote{The SP algorithm, since it looks at all permutations of the variables, is exponential and so scales in Tetrad comfortably only to about 11 variables (we have 10 here). Still, the implementation in Tetrad allows up to 11 variables per knowledge tier, so we are able to extend its functionality to more variables. SP is being substituted in SP-FCI for FGES in the GFCI algorithm.}








\begin{verbatim}
import pandas as pd
import graphviz as gviz

import jpype
import jpype.imports
jpype.startJVM(classpath=[f"resources/tetrad-gui-current-launch.jar"])

import pytetrad.tools.translate as ptt
import pytetrad.tools.visualize as ptv
import edu.cmu.tetrad.search as ts
import edu.cmu.tetrad.data as td


tiers = [['age', 'gender', 'height', 'weight', 'resting_heart', 'device', 'activity'],
         ['steps', 'heart_rate', 'calories', 'distance']]

df = pd.read_csv("resources/aw-fb-pruned18.data.mixed.numeric.txt", sep="\t")
df = df[tiers[0] + tiers[1]]
df = df.astype({col: int for col in ["gender", "device", "activity"]})

knowledge = td.Knowledge()
knowledge.setTierForbiddenWithin(0, True)
for col in tiers[0]: knowledge.addToTier(0, col)
for col in tiers[1]: knowledge.addToTier(1, col)
\end{verbatim}

\begin{figure}
\floatconts
  {fig: image}
  {\caption{Graph for the JPype example in the text for "Apple Watch Fitbit."}}
  {\includegraphics[width=0.6\linewidth]{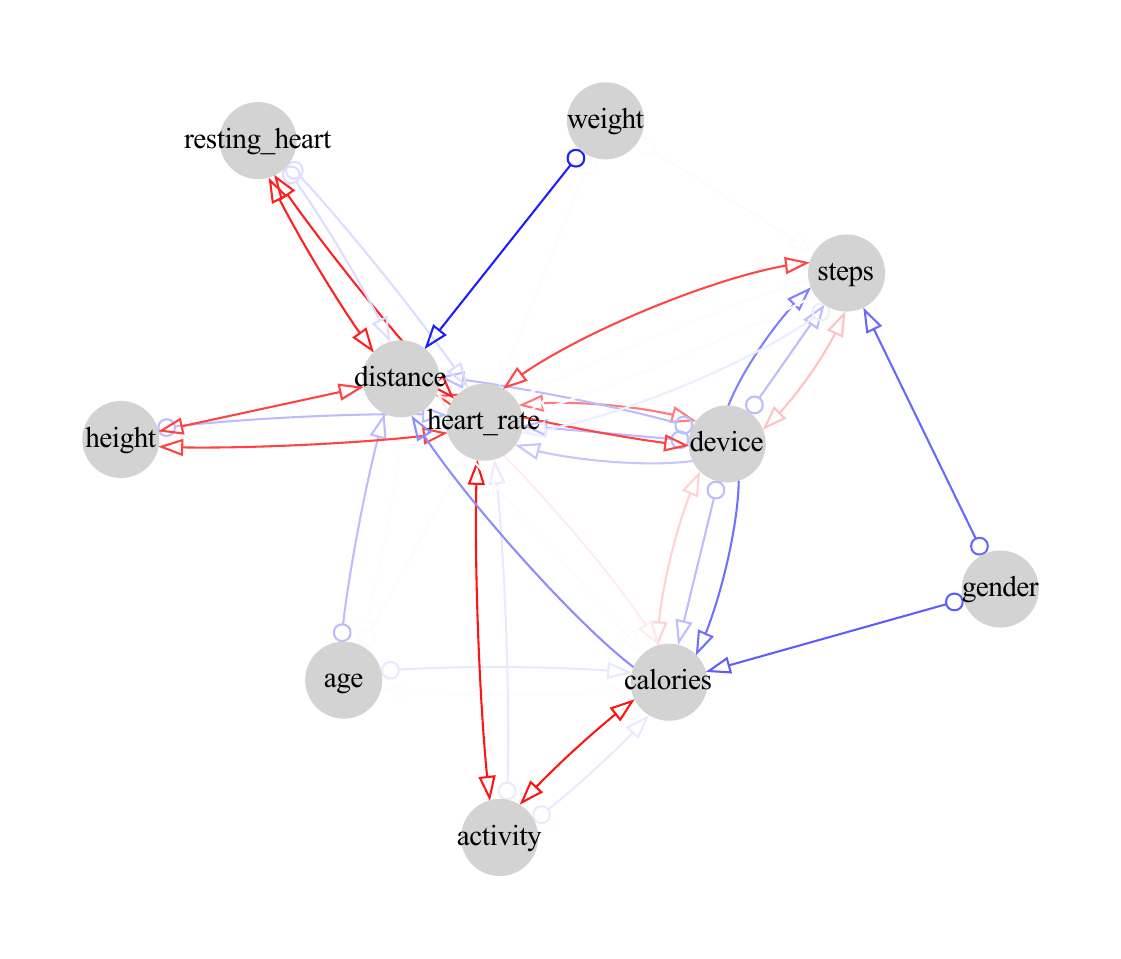}}
  \label{jpype_plot}
\end{figure}

Now we bootstrap and record the results. Note that since some columns are discrete and some continuous, we need to use a score suitable for mixed data types. Finally, we visualize and print the results. The bootstrapped results are shown in Table \ref{tab:my_label}; the graph plot is shown in Figure \ref{jpype_plot}.

\begin{verbatim}
reps = 10
graphs = []
for rep in range(reps):
    data = ptt.pandas_data_to_tetrad(df.sample(frac=1, replace=True))

    score = ts.score.DegenerateGaussianScore(data, True)
    score.setPenaltyDiscount(2)
    test = ts.test.ScoreIndTest(score, data)

    alg = ts.SpFci(test, score)
    alg.setKnowledge(knowledge)
    graphs.append(alg.search())

probs = ptv.graphs_to_probs(graphs)
graph_attr = {"viewport": "600", "outputorder": "edgesfirst"}
gdot = gviz.Graph(format="pdf", engine="neato", graph_attr=graph_attr)

gdot = ptv.write_gdot(gdot, probs, length=2)
gdot.render(filename="apple_fitbit", cleanup=True, quiet=True)
gdot.clear()
\end{verbatim}

\begin{table}[ht]
    \centering
    \begin{tabular}{lcccccc}
        Adjacency & $\leftrightarrow$ & $\circ\!\!\rightarrow$ & $\rightarrow$ & $\leftarrow\!\!\circ$ & $\leftarrow$ \\
        \hline
        (activity, calories) & 0.93 & 0.07 & - & - & - \\
        (activity, heart rate) & 0.93 & 0.07 & - & - & - \\
        (calories, device) & 0.22 & - & - & 0.30 & 0.48 \\
        (device, distance) & 0.70 & 0.30 & - & - & - \\
        (device, heart rate) & - & 0.45 & - & 0.30 & 0.25 \\
        (device, steps) & - & 0.19 & 0.51 & 0.30 & - \\
        (distance, resting heart) & 0.80 & - & - & 0.20 & - \\
        (heart rate, height) & 0.71 & - & - & 0.29 & - \\
        (heart rate, resting heart) & 0.80 & - & - & 0.20 & - \\
        (distance, weight) & - & - & - & 0.86 & - \\
        (distance, height) & 0.71 & - & - & 0.03 & - \\
        (heart rate, steps) & - & 0.78 & - & 0.02 & 0.01 \\
        (gender, steps) & - & 0.72 & - & - & - \\
        (calories, gender) & - & - & - & 0.71 & - \\
        (calories, distance) & 0.06 & - & 0.47 & - & - \\
        (age, distance) & - & 0.29 & - & - & - \\
        (age, calories) & - & 0.05 & - & - & - \\
        (age, heart rate) & - & 0.01 & - & - & - \\
        (heart rate, weight) & - & - & - & 0.01 & - \\
    \end{tabular}
    \caption{Results of a 100-fold bootstrapping of the Apple Watch Fitbit data, as described in the text. Frequencies for each edge type encountered over 100 folds are given.}
    \label{tab:my_label}
\end{table}

\section{A Code Example in R}
\label{sec:example-r}

We provide support for R via py-tetrad and the R project, Reticulate, in a project we call ``rpy-tetrad,'' located in the 'R' subdirectory of the py-tetrad project. Instructions for setting up the project to run in R or RStudio; these instructions, and rpy-tetrad itself, have been tested and shown to work on Mac, Linux, and Windows. Note that the interface in R to Tetrad is limited to what is available in the TetradSearch class in py-tetrad, so if more methods are needed in R, more methods must be added to the py-tetrad class so R can access them; this is straightforward. One cannot use JPype commands directly from R; these must be routed through Python.

We give an example in this section; more examples are available in the 'R' subdirectory of py-tetrad. We use data from a NASA Airfoil Self-Noise experiment \citep{misc_airfoil_self-noise_291} with six variables. In this example, the data loader in R loads the data as a mixture of continuous and discrete (integer) variables, which the converter in py-tetrad will translate as a mixture of continuous data in Tetrad. To enforce the constraint that all variables are interpreted as continuous, we need first to tell R that the columns in this data frame are all to be interpreted as 'numeric'; a line in the code does this. Then a test defined over continuous variables ('Fisher Z') and a score defined over continuous variables ("SEM BIC") can be used to run the search algorithm. TetradSearch will report if the search, test, or score is for one data type but another data type is given.

As in the Python example, we here appeal to background knowledge. We use the feature in the TetradSearch class that allows us to specify background knowledge as `temporal tiers,' where variables in later tiers are known not to cause variables in earlier tiers. For this example, we know that 'pressure' (sound pressure) in this NASA experiment cannot cause any other experimental or airfoil design variables, so we put it in a later tier. (One may also forbid or require particular edges as part of background knowledge.) If we use such background knowledge in a search, we expect the algorithm to honor it. Not all algorithms in Tetrad are designed to honor background knowledge; if one chooses an algorithm that does not and provides background knowledge, an error will be reported. Examples of algorithms that are not currently designed to honor background knowledge include ICA LiNGAM, Direct LiNGAM, and CCD, though this may change.

\begin{figure}
\floatconts
  {fig:image}
  {\caption{Plot matrix for the R code "NASA Airfoil Self-Noise" example; this is displayed in the Plots window in RStudio. The psych package in R produced this plot matrix.}}
  {\includegraphics[width=0.5\linewidth]{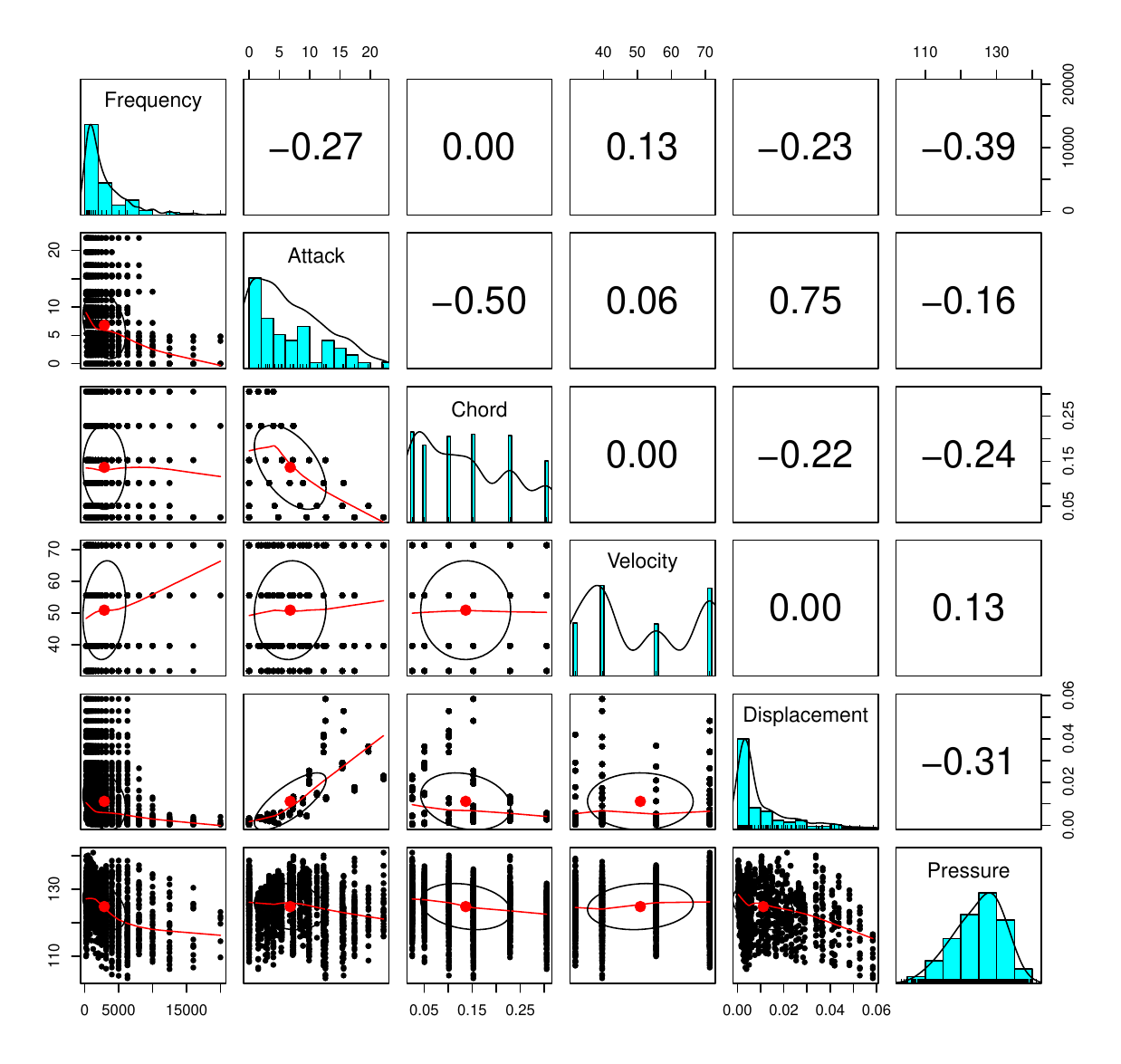}}
  \label{r_example_plot_matrix}
\end{figure}

\begin{figure}
\floatconts
  {fig:image}
  {\caption{Plot of the graph for the R code "NASA Airfoil Self-Noise" example; this is displayed in the Viewer window in RStudio. This plot is produced by Graphviz using the DiagrammaR package in R.}}
  {\includegraphics[width=0.4\linewidth]{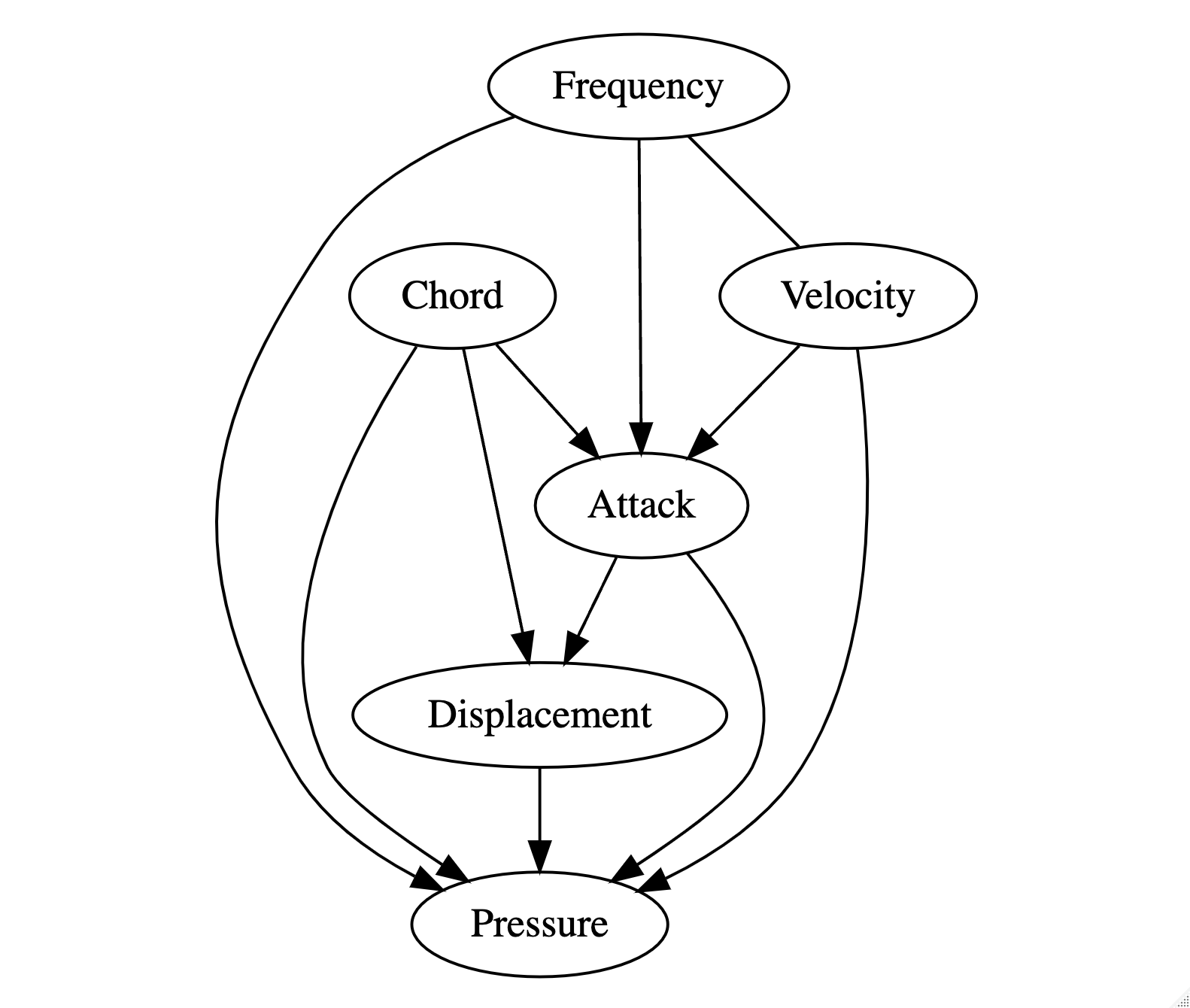}}
  \label{r_example_plot}
\end{figure}

This example is written to work in RStudio, so that a histogram/scatterplot plot matrix (Figure \ref{r_example_plot_matrix}) is displayed in the Plots window, and the plot of the graph returned by the algorithm is displayed in the Viewer window (e.g., Figure \ref{r_example_plot}). We have included this code example in the py-tetrad repository.\footnote{\href{https://github.com/cmu-phil/py-tetrad/blob/main/pytetrad/R/sample_r_code7.R}{https://github.com/cmu-phil/py-tetrad/blob/main/pytetrad/R/sample\_r\_code7.R}}

\begin{figure}[ht]
\floatconts
  {fig:image}
  {\caption{Plot of the 30-fold bootstrapping graph for the R code "NASA Airfoil Self-Noise" example displayed in the Viewer window in RStudio. This plot is produced by Graphviz using the DiagrammaR package in R.}}
  {\includegraphics[width=0.8\linewidth]{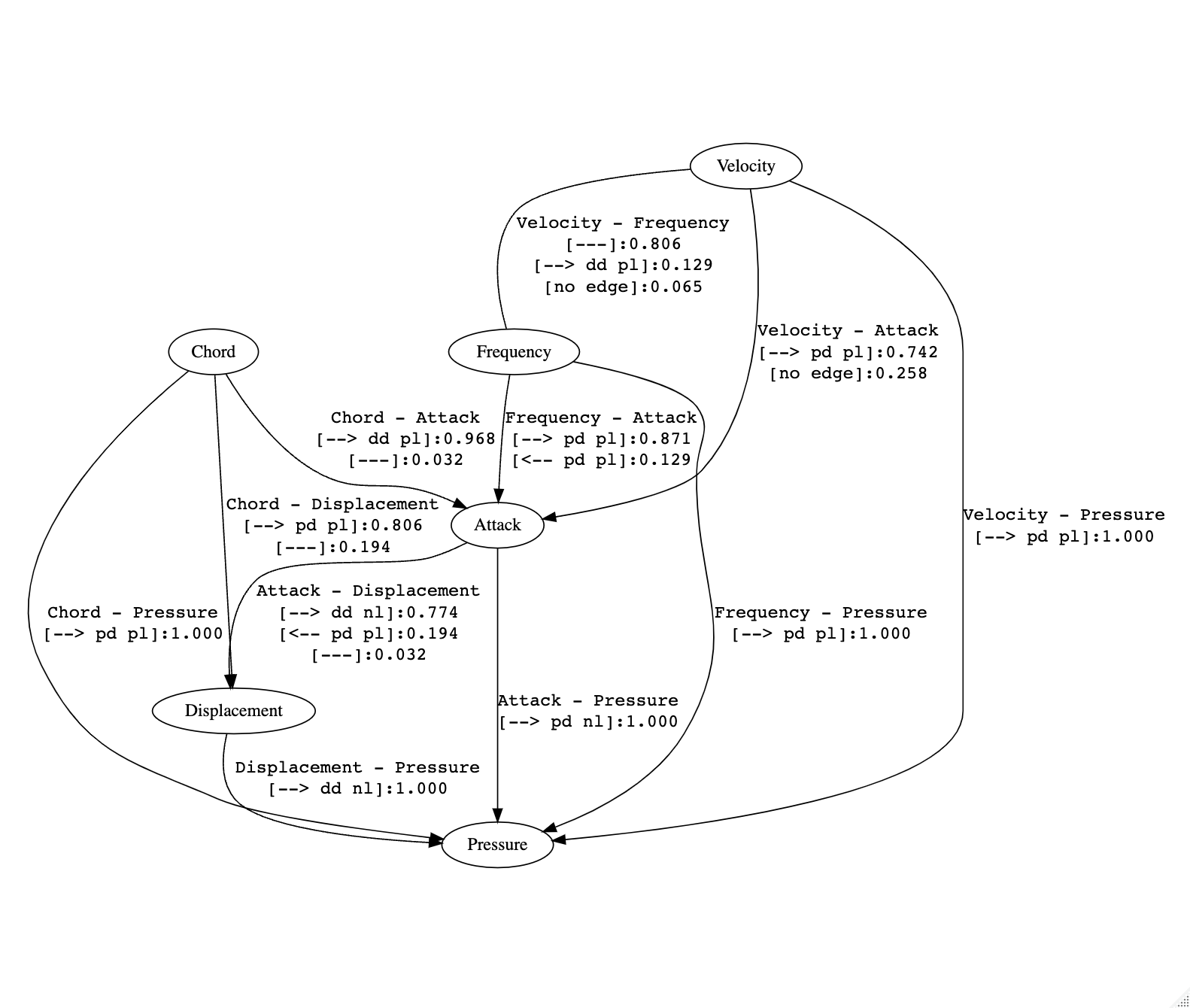}}
  \label{r_example_bootstrapping}
\end{figure}

Tetrad also has a bootstrapping facility that TetradSearch makes available if one prefers to use it. (As in the Python example, the user may prefer to write their own bootstrapping code.) If automatic bootstrapping is used, a graph is plotted that shows frequencies of occurrence of each edge observed in any bootstrapping models. We show bootstrapping results with 30 folds (Figure \ref{r_example_bootstrapping}).\footnote{It should be noted that the TetradSearch class can also be used in Python as well for those who don't wish to make JPype calls themselves. Examples of this are provided in the py-tetrad package.} We use here the GRaSP algorithm, a permutation algorithm that relaxes the faithfulness assumption, with the linear Gaussian BIC score.

\begin{verbatim}
setwd("~/py-tetrad/pytetrad")
library(reticulate)

data <- read.table("./resources/airfoil-self-noise.continuous.txt",
                   header=TRUE)
i <- c(1, 6)
data[ , i] <- apply(data[ , i], 2, function(x) as.numeric(x))

library(psych)
pairs.panels(data, method = "pearson") 

source_python("tools/TetradSearch.py")
ts <- TetradSearch(data)

ts$add_to_tier(1, "Attack")
ts$add_to_tier(1, "Chord")
ts$add_to_tier(1, "Velocity")
ts$add_to_tier(1, "Displacement")
ts$add_to_tier(1, "Frequency")
ts$add_to_tier(2, "Pressure")

ts$use_sem_bic(penalty_discount=2)
ts$use_fisher_z()
ts$run_grasp()

print(ts$get_string())

library(DiagrammeR)
dot <- ts$get_dot()
grViz(dot)

ts$set_bootstrapping(numberResampling = 30)
ts$run_grasp()
dot <- ts$get_dot()
grViz(dot)
\end{verbatim}

\section{Conclusion}
\label{sec:conclusion}

In Python, the JPype package allows access to arbitrary code in Tetrad. In both Tetrad and R, the TetradSearch class allows users to access Tetrad's most commonly used functionality without directly using JPype. If one wishes to do JPype programming, the entire codebase of Tetrad becomes available for Python scripts. These Python or R scripts can be published, shared, and reused. Despite its performance advantages over Python in many areas, Java is not a good scripting language, so publishing Java classes as scripts is a bit forced. Also, accessing other languages from Java is tricky, whereas accessing Java from Python with JPype is not, as we have shown. So py-tetrad scripts can easily take cognizance of the entire functionality available in Python, including functionality from Java.

One known issue is that the installation process can be a bit cumbersome. We hope to simplify this process so users can install py-tetrad using pip and rpy-tetrad using CRAN. Also, initial responses to these tools have been overwhelmingly positive, but feedback is welcome for issues encountered in py-tetrad or rpy-tetrad. Installing Grasphviz in Python is also challenging but must be left as an exercise for the reader. Suggestions for new features to include in py-tetrad, rpy-tetrad, or Tetrad are welcome.\footnote{Bug reports and feature suggestions may be sent to us by email or, preferably, submitted to our GitHub Issue Tracker at \href{https://github.com/cmu-phil/pytetrad/issues}{https://github.com/cmu-phil/pytetrad/issues}.}

\section{Citations and Bibliography}
\label{sec:cite}

\acks{We thank our anonymous reviewers for their detailed comments. We thank Yasuhiro Shimodaira and Kelvin Lim for their feedback, especially with our R implementation, and for ensuring our projects run well on Windows. We also thank Peter Spirtes for encouraging this project and setting the goal of breaking down barriers between projects. We especially thank the authors of JPype and Reticulate for their excellent tools. Ramsey's work on this project was partly funded and supported by the Department of Defense under Contract No. FA8702-15-D-0002 with Carnegie Mellon University for the operation of the Software Engineering Institute, a federally funded research and development center. Andrews' work on this project was funded and supported by the Comorbidity: Substance Use Disorders and Other Psychiatric Conditions Training Program T32DA037183.}

\bibliography{refs}

\end{document}